\documentclass[prb,aps,twocolumn]{revtex4-1}
\pdfoutput=1
\usepackage{dcolumn}
\usepackage{bm}

\usepackage[colorlinks,hyperindex, urlcolor=blue, linkcolor=blue,citecolor=black, linkbordercolor={.7 .8 .8}]{hyperref}
\usepackage{graphicx}
\usepackage{amsfonts}
\usepackage{amsmath}
\usepackage{amssymb}
\usepackage{amsbsy}
\usepackage{nicefrac}

\newcommand{\sxx}{$\sigma_{xx}$}
\newcommand{\drxx}{$\Delta \rho_{xx}/\rho_{xx}$}

\providecommand{\abs}[1]{\lvert#1\rvert}

\begin{document}

\title {Transport in indium-decorated graphene}

\author{U. Chandni}
\affiliation{Institute of Quantum Information and Matter, Department of Physics, California Institute of Technology, 1200 E. California Blvd., Pasadena, California 91125, USA}
\author{Erik A. Henriksen}
\altaffiliation[Now at ]{Department of Physics, Washington University in St. Louis, St. Louis, MO 63130, USA}
\email{henriksen@wustl.edu}
\affiliation{Institute of Quantum Information and Matter, Department of Physics, California Institute of Technology, 1200 E. California Blvd., Pasadena, California 91125, USA}
\author{J.P. Eisenstein}
\affiliation{Institute of Quantum Information and Matter, Department of Physics, California Institute of Technology, 1200 E. California Blvd., Pasadena, California 91125, USA}

\begin{abstract}
The electronic transport properties of single layer graphene having a dilute coating of indium adatoms has been investigated. Our studies establish that isolated indium atoms donate electrons to graphene and become a source of charged impurity scattering, affecting the conductivity as well as magnetotransport properties of the pristine graphene. Notably, a positive magnetoresistance is observed over a wide density range after In doping. The low field magnetoresistance carries signatures of quantum interference effects which are significantly altered by the adatoms. 
\end{abstract}

\date{\today}

\maketitle

\section{Introduction}

Surface adsorbates have the potential to open novel avenues for tailoring the electronic, optical, magnetic and chemical properties of many material systems. In addition to influencing the electronic scattering mechanisms, adatoms can couple to the spin, orbital and charge degrees of freedom and even bestow their own distinctive properties upon the substrate material \cite{bergmann82}. These effects are enhanced in thin films due to the increased surface to volume ratio. Graphene, a single-atom-thick sheet of carbon atoms behaving as a zero band-gap semimetal with a linear Dirac spectrum, is therefore a compelling host in this respect. While experiments on graphene have uncovered a myriad of intriguing properties to date \cite{dassarma11}, the effects of adatoms on the electron transport in graphene have only begun to be explored. Early studies on potassium doped graphene were instrumental in demonstrating the role of charged impurity scattering on the conductivity\cite{chen07}. The adsorption of elements such as hydrogen \cite{elias09,balakrishnan13}, oxygen \cite{ito08} and fluorine \cite{hong12} have been found to strongly impact the electronic behavior, inducing insulating band gaps or local magnetic moments. Transition metal adatoms on graphene are of particular interest due to a spate of intriguing theoretical predictions. Several $5d$ atoms are expected to induce novel topological behavior such as quantum spin Hall or quantum anomalous Hall effects \cite{weeks11,zhang12,hu12}. Meanwhile the expectation of magnetic moments arising from $3d$ metal adatoms has received experimental support \cite{eelbo13}. Thus further studies into such heavy adatoms are timely and essential for engineering newer graphene devices. 

In the present work, we employ indium (In) to investigate the influence of heavy adatoms on the electrical transport of graphene. We find that dilute In coverages (less than 1\%) on SiO$_2$-supported graphene significantly charge dope the system, leading to increased charged impurity scattering and decreased carrier mobilities.  At the same time, the magneto-resistance at the few-Tesla scale reveal signatures of an In-induced enhancement of the charge density inhomogeneities (commonly referred to as ``puddles'') around the Dirac point.  At low magnetic fields ($\lesssim 50$ mT) clear signatures of weak localization~\cite{tikhonenko08,tikhonenko09} and universal conductance fluctuations are seen, with the In adatoms reducing the amplitude and expanding the width of the zero-field weak localization anomaly and suppressing the conductance fluctuations. 

\section{Experimental}
Measurements of the graphene samples were performed {\it in situ} before and after the controlled deposition of In atoms in a custom-built cryostat inserted into a liquid helium dewar. The sample stage, held within the cryogenically-established ultra-high vacuum (UHV) environment, may be controllably heated to initially desorb surface contaminants and later to reverse the charge doping created by the indium adatoms.  A thin, indium-coated tungsten wire, located about 15 cm below the sample stage, allows for deposition of a dilute quench-condensed film of In onto the cold graphene sample.  The average concentration of deposited In adatoms may be estimated from the observed changes in the graphene transport due to charge doping by the indium; theoretical estimates\cite{weeks11,chan08} suggest roughly 0.8 electron donated per In atom.  The deposition rate is highly controllable via the current passed through the tungsten wire source. The charge doping induced by indium deposition can be reversed upon annealing the sample to temperatures above 450 K, with the sample returning close to its original state. This behavior was reproducible in several samples. Apart from the deposition and annealing processes, the devices are stable at low temperature for periods of several weeks, due to the intrinsically high vacuum of the cryogenic environment. Each device consists of graphene mechanically exfoliated from Kish graphite onto a Si/SiO$_2$ wafer having an oxide thickness of 285 nm. Voltages $V_g$ applied to the degenerately-doped substrate control the graphene free carrier density, $n \propto (V_g-V_0)$, with the offset voltage $V_0\sim 15-20$ V determined by charged impurities in the SiO$_2$. Electrical contacts were made using standard electron beam lithography. The samples were subsequently etched in an oxygen plasma to form a Hall bar and rinsed clean with solvents. Here we report results from a single Hall bar device having wide and narrow central regions of width $W =8$ $\mu$m and 1 $\mu$m respectively, see Figure 1(a). 

\begin{figure}[t]
\includegraphics[width=1 \columnwidth]{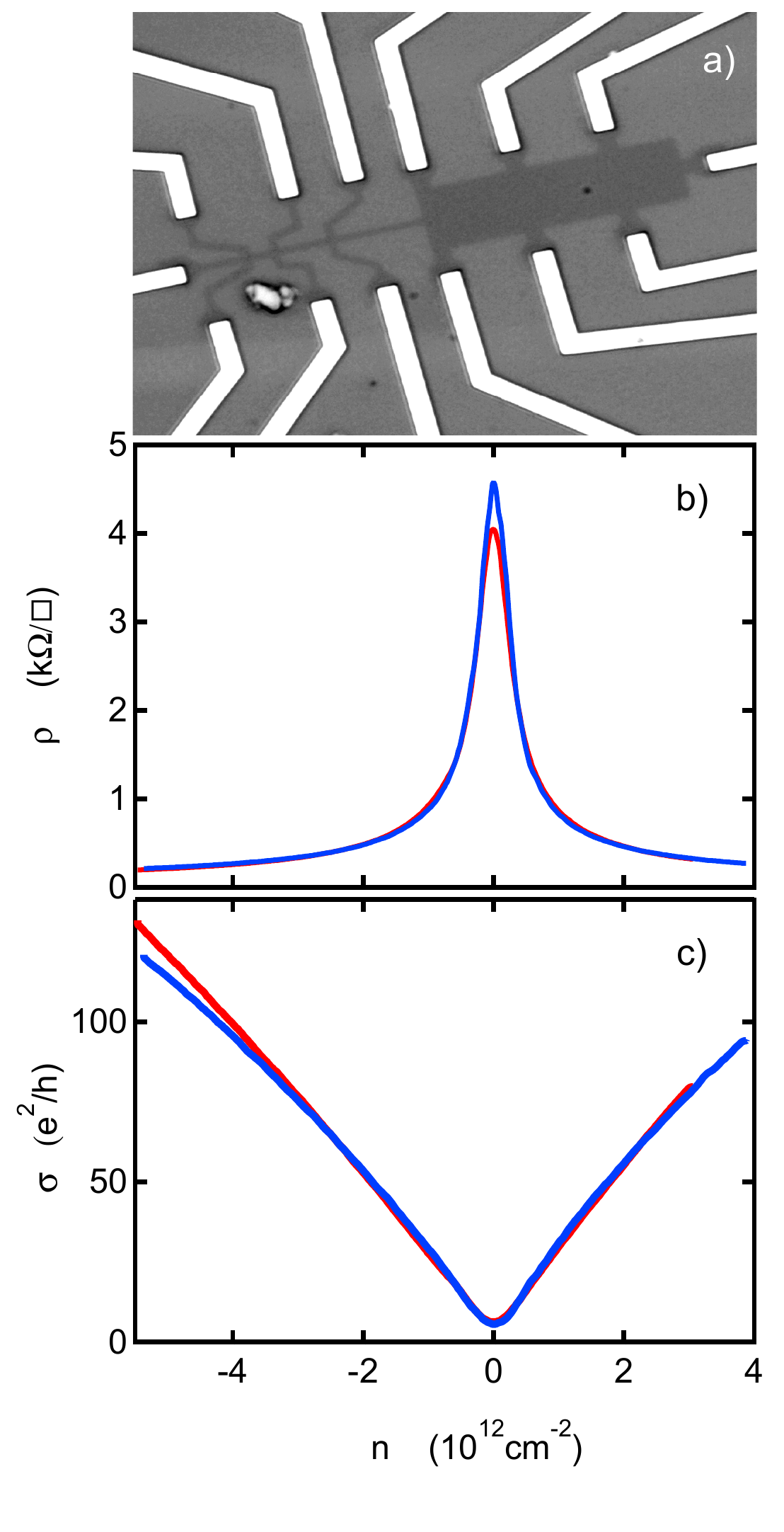}
\caption{(Color online) (a) Contrast-enhanced image of single-layer graphene Hall bar device. The Hall bar has both a wide (8 $\mu$m) region and a narrow (1 $\mu$m) region.  (b) Resistivity $\rho_{xx}$ vs. density $n$ at $T = 12$ K, for the wide (red) and narrow (blue) regions of the graphene device prior to deposition with In adatoms.  c) Plot of the conductivity $\sigma_{xx}$ vs. $n$ plot clearly reveals a weak sub-linear contribution to the density dependence of $\sigma_{xx}$ in the as-made device.} 
\end{figure}

\section{Results}
\subsection{Transport at zero magnetic field}
Figure 1 (b) shows the longitudinal resistivity $\rho_{xx}$ (in ohms per square) and conductivity $\sigma_{xx}$ (in units of $e^2/h$) of both the wide (red) and narrow (blue) regions of the as-made (i.e. before In deposition) device as functions of the free carrier density $n$, at $T = 12$ K.  (The gate voltage versus density calibration is established via observations of Shubnikov-de Haas magneto-oscillations in $\rho_{xx}$.)  The resistivity peak appears extremely similar in the two regions, differing noticeably only at the Dirac point.  The conductivity is roughly linear in density away from the Dirac point, with the slope $d\sigma_{xx}/dn$ implying an average mobility of $\mu \approx 6000$ cm$^2$/Vs.  As commonly observed, there is also a small sub-linear contribution to $\sigma_{xx}(n)$.  The sub-linear contribution appears to be stronger in the narrow region of the device, at least at high densities on the hole-side of the Dirac point.  While the dominant linear density dependence of $\sigma_{xx}$ is believed to arise from long-range charged impurity scattering, the sub-linear contribution is generally attributed to short-range (on the lattice scale) scatterers~\cite{adam07}.  The edge of the graphene sheet is one obvious source of short-range scattering and this might explain the enhanced sub-linearity of the conductivity we observe in the narrow region of the device.  

\begin{figure}[t]
\includegraphics[width=1 \columnwidth]{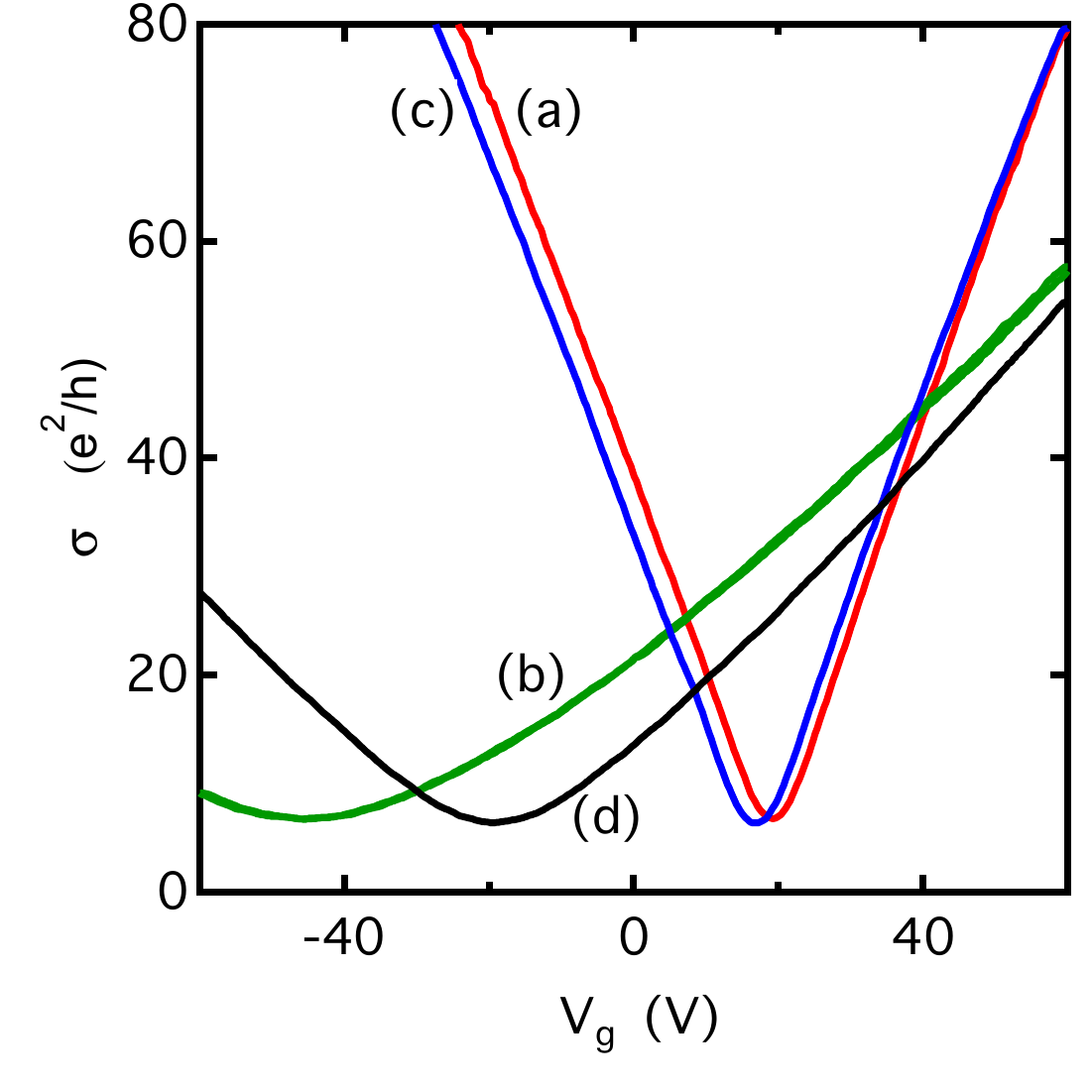}
\caption{(Color online) Effect of In deposition on graphene conductivity $\sigma_{xx}$ vs. gate voltage $V_g$ at $T = 12$ K.  Data is from the wide region of the Hall bar, with trace (a) taken before any In deposition, trace (b) after the first In deposition, trace (c) after warming the sample to 450 K, and finally trace (d) after a second In deposition.} 
\end{figure}
The charge doping and associated changes in electronic transport at zero magnetic field due to In deposition are demonstrated in Fig. 2 where the conductivity \sxx\ is plotted versus gate voltage, $V_g$.  The data shown is from the wide region of the device; the results from the narrow region are almost identical.   Four traces, (a) through (d), are shown in the figure.  Trace (a) was acquired after the graphene had been annealed at 200$^\circ$ C in the cryogenically established UHV environment, but prior to any heating of the In evaporation source. The Dirac minimum conductivity point lies near $V_g = 20$ V and the steep, nearly linear, increase in \sxx\ away from this point correspond to average electron and hole mobilities of $\mu \gtrsim 6000$ cm$^2$/Vs, as mentioned above.  Subsequent to recording trace (a), the evaporation source was heated and In atoms began to adsorb onto the graphene surface.  This process was monitored in real-time by observing the graphene conductivity.  Trace (b), recorded after the evaporation was stopped, immediately reveals several qualitative effects of the In deposition: First, the Dirac point has shifted to more negative $V_g$, demonstrating electron doping of the graphene by the In adatoms.  From the magnitude of the shift, trace (b) corresponds to a net electron doping of about $4.5 \times 10^{12}$ cm$^{-2}$.   Second, the slope $d\sigma_{xx}/dV_g$ away from the Dirac point is much less than in trace a), indicating significantly reduced carrier mobility.  Third, the curvature $d^2\sigma_{xx}/dV^2_g$ around the Dirac point is substantially smaller than in trace (a).  This suggests that the In deposition has enhanced the carrier density inhomogeneity at the Dirac point.  Finally, though not readily apparent in Fig. 2, the weak sub-linearity of \sxx\ vs. density apparent in Fig. 1 (c) is no longer observed after In has been deposited; in fact, away from the Dirac point, $\sigma_{xx}(V_g)$ shows a small $super$-linearity after In deposition.  These various effects are broadly similar to those observed in experiments on graphene doped with potassium or other adatoms \cite{chen07,adam07,yan11,pi09,dassarma11}, and may reasonably be expected if indium atoms donate electrons to graphene and become charged impurity scattering centers.

After trace (b) was recorded, the graphene sample was briefly heated $in~situ$ to about 450 K.  After re-cooling to $T = 12$ K, trace (c) was recorded.  As the figure makes clear, the $\sigma_{xx}(V_g)$ characteristic of the graphene sample is almost exactly the same as it was before any In was deposited; there is only a small shift in the location of the Dirac point.  Heating to 450 K is apparently sufficient to `clean' the graphene sample.  Whether this cleaning occurs because the In adatoms have desorbed, migrated away from the conducting region of the sample, or have been rendered benign in some other way, is at present unknown.  

Finally, a second, briefer, In deposition was made, resulting in \sxx\ trace (d).  As intended, the doping is less than for trace (b), and corresponds to about $2.5 \times 10^{12}$ cm$^{-2}$.  All of the effects observed in trace (b) are also seen in trace (d), only now, as expected, less intensely.

At low temperatures, the conductivity \sxx\ of graphene, away from the Dirac point, is often assumed to reflect two distinct sources of scattering, screened charged impurities and abrupt short-range scatterers (edges, lattice defects, etc.)\cite{nomura06,adam07}:  
\begin{equation}
\sigma_{xx}^{-1}(n)=\sigma^{-1}_{CI}(n)+\sigma^{-1}_{SR}~.
\end{equation}
For charged impurities a distance $d<<1$ nm away from the graphene plane, $\sigma_{CI}(n)$ is proportional to the carrier density $n$:  $\sigma_{CI}(n) = C~ \abs{n}/n_{imp}$, where $n_{imp}$ is the impurity density and $C$ is a constant estimated \cite{adam07} to be $C = 20 e^2/h$ in the limit $d \rightarrow 0$.  (At finite $d$, Adam {\it et al.} find $\sigma_{CI}(n)$ to be super-linear in $n$, increasingly so as $d$ rises\cite{adam07}.)  In contrast to this, the short-range contribution to the conductivity, $\sigma_{SR}$, is expected to be independent of density.  

For our sample, the conductivity $\sigma_{xx}$, away from the Dirac point, is nearly linear in density both before and after the deposition of indium.  This suggests that nearby screened charged impurities dominate the carrier mobility $\mu = \sigma_{xx}/\abs{n}e$.  By making the reasonable assumption that the number of additional charged impurities, $\Delta n_{imp}$, due to the indium deposition equals the charge doping deduced from the shift of the Dirac peak, we can estimate the distance $d$ between the adatoms and the graphene plane by comparing the conductivity before and after the deposition and using the theory of Adam {\it et al.}\cite{adam07}.  For carrier densities between $n = 3$ and $4 \times 10^{12}$ cm$^{-2}$, we find $d \approx 0.8 - 0.9$ nm for both the first and second In deposition.  At these $d$ values, the theory shows somewhat more super-linearity in $\sigma_{xx}(n)$ than we observe; this is plausibly the result of short-range scatterers present in our sample but not included in the theory.  We note that recent density functional theory (DFT) calculations suggest that chemisorbed In atoms reside much closer, favoring positions about $d\approx 0.24$ nm above the center of the hexagons of carbon atoms in the graphene\cite{aliceapc,nakada11}.

\subsection{Magneto-resistance at intermediate magnetic fields}
We turn now to the impact of the indium adatoms on the resistivity coefficients (Hall, $\rho_{xy}$, and longitudinal, $\rho_{xx}$) in a magnetic field $B$. We concentrate first on fields $B \gtrsim 50$ mT in order to avoid the quantum interference effects which we discuss subsequently.  

\begin{figure}
\includegraphics[width=1 \columnwidth]{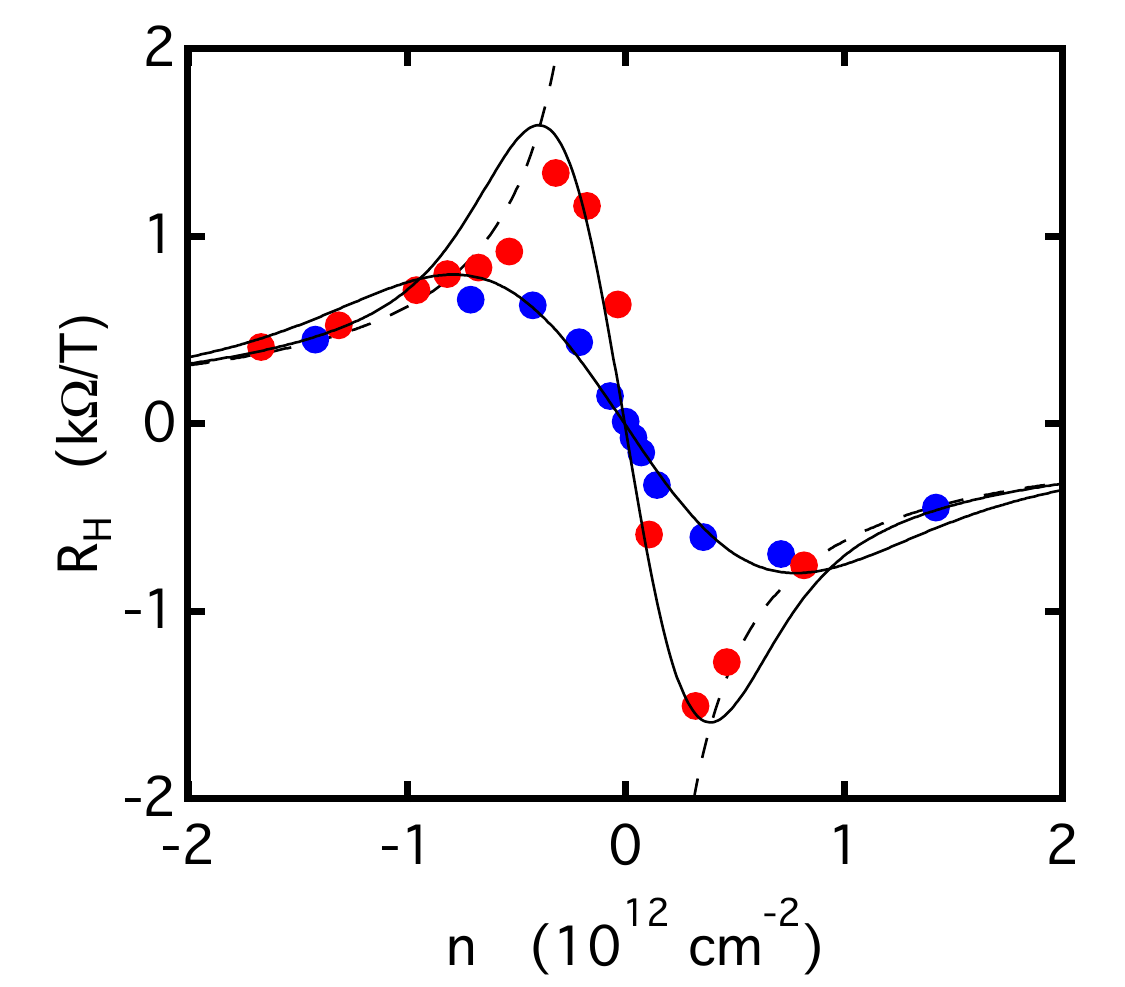}
\caption{(Color online) Hall coefficient vs. density near the Dirac point.  The red and blue data points correspond to $R_H$ measurements before and after the deposition of about $2.5 \times 10^{12}$ cm$^{-2}$ indium adatoms, respectively.  The dashed curve shows the ideal $R_H=-1/ne$ behavior while the two light solid lines display convolutions of the ideal behavior with gaussian density distributions of rms width $\sigma_n = 0.3$ and $0.6 \times 10^{12}$ cm$^{-2}$.} 
\end{figure}

Figure 3 shows the Hall coefficient $R_H \equiv d\rho_{xy}/dB$, measured in the wide region of the device at $T=12$ K, as a function of free carrier density $n$, where $n$ is computed from the gate voltage (relative to the Dirac point) and the known capacitance between the graphene and the conducting Si substrate.  Two data sets are shown: The red data points were obtained from the graphene sample before any indium deposition, while the blue data were obtained after about $2.5 \times 10^{12}$ cm$^{-2}$ indium atoms were deposited.  Owing to charge density inhomogeneity, the Hall coefficient passes smoothly through $R_H=0$ at $n=0$ in both cases. It is clear from the figure that the smearing of the divergence at $n=0$ of the ideal $R_H=-1/ne$ Hall coefficient (indicated by the dashed solid line in the figure) is significantly stronger when the indium is present.  The two light solid lines are simple convolutions of the ideal $R_H$ behavior with gaussian density distributions of rms widths $\sigma_n = 0.3$ and $0.6 \times 10^{12}$ cm$^{-2}$; these roughly approximate the observed behavior of $R_H(n)$ for the clean and indium-decorated graphene sample, respectively.  

The data in Fig. 3 corroborate our earlier conclusion that the deposition of indium adatoms increases the charge density inhomogeneity $\sigma_n$ near the graphene Dirac point.  The theory of Adam {\it et al.}\cite{adam07} suggests that a sheet of $2.5 \times 10^{12}$ cm$^{-2}$ charged impurities positioned $d = 0.9$ nm away from graphene will induce charge density fluctuations of rms amplitude $\sigma_n = 0.6 \times 10^{12}$ cm$^{-2}$ around the Dirac point.  This remarkable agreement with the blue data set in Fig. 3 is somewhat misleading since, as the red data set in the figure proves, density fluctuations  ($\sigma_n \approx 0.3 \times 10^{12}$ cm$^{-2}$) are present in our sample even prior to the indium deposition.  However, since it is reasonable to assume that these prior density fluctuations are statistically independent of the fluctuations induced by the In adatoms, the two sources of inhomogeneity would add in quadrature to produce $\sigma_{n,tot} \approx 0.7 \times 10^{12}$ cm$^{-2}$, still in good agreement with the blue data set in Fig. 3.   We note in passing that if $d\approx 0.25$ nm, as DFT suggests\cite{aliceapc}, the theory of Adam {\it et al.}\cite{adam07} would predict $\sigma_n \approx 1.1 \times 10^{12}$ cm$^{-2}$.  Our data is not consistent with such a large density inhomogeneity.

Away from the Dirac point, at densities $|n| \gtrsim 10^{12}$ cm$^{-2}$, the longitudinal resistivity $\rho_{xx}$  of the sample exhibits very little magnetic field dependence for $B \lesssim 3$ T.  This is illustrated in Fig. 4(a) where the fractional change in the resistivity $\Delta \rho_{xx} /\rho_{xx} \equiv (\rho_{xx}(B)-\rho_{xx}(0))/\rho_{xx}(0)$ measured in the wide region of the device is plotted versus magnetic field at the hole-like density of $n=-2.8 \times 10^{12}$ cm$^{-2}$.  Again two data sets are shown, one with and one without approximately $2.5 \times 10^{12}$ cm$^{-2}$ In adatoms present.  This very weak magnetic field dependence is consistent with the behavior of a simple Drude metal.  (Beyond $B\approx 3$ T clear Shubnikov-de Haas oscillations emerge in $\rho_{xx}$ data from the clean graphene sample.  Only weak hints of oscillations are seen, at the highest magnetic fields, in the lower mobility, In-decorated sample.)

As the $n=0$ Dirac point is approached, this very weak magneto-resistance is replaced by a strong positive magnetoresistance.  This is  
shown in Figs. 4(b) and 4(c) where \drxx\ data from the wide region of the clean and In-decorated graphene sample, respectively, are displayed at various free carrier densities, $n$.  At $n=0$ (red traces) both the clean and In-decorated sample exhibit a nearly linear magnetoresistance, with $\rho_{xx}$ increasing by about 70 (50) \% in the clean (In-decorated) graphene sample by $B = 2$ T.

\begin{figure}
\includegraphics[width=1 \columnwidth]{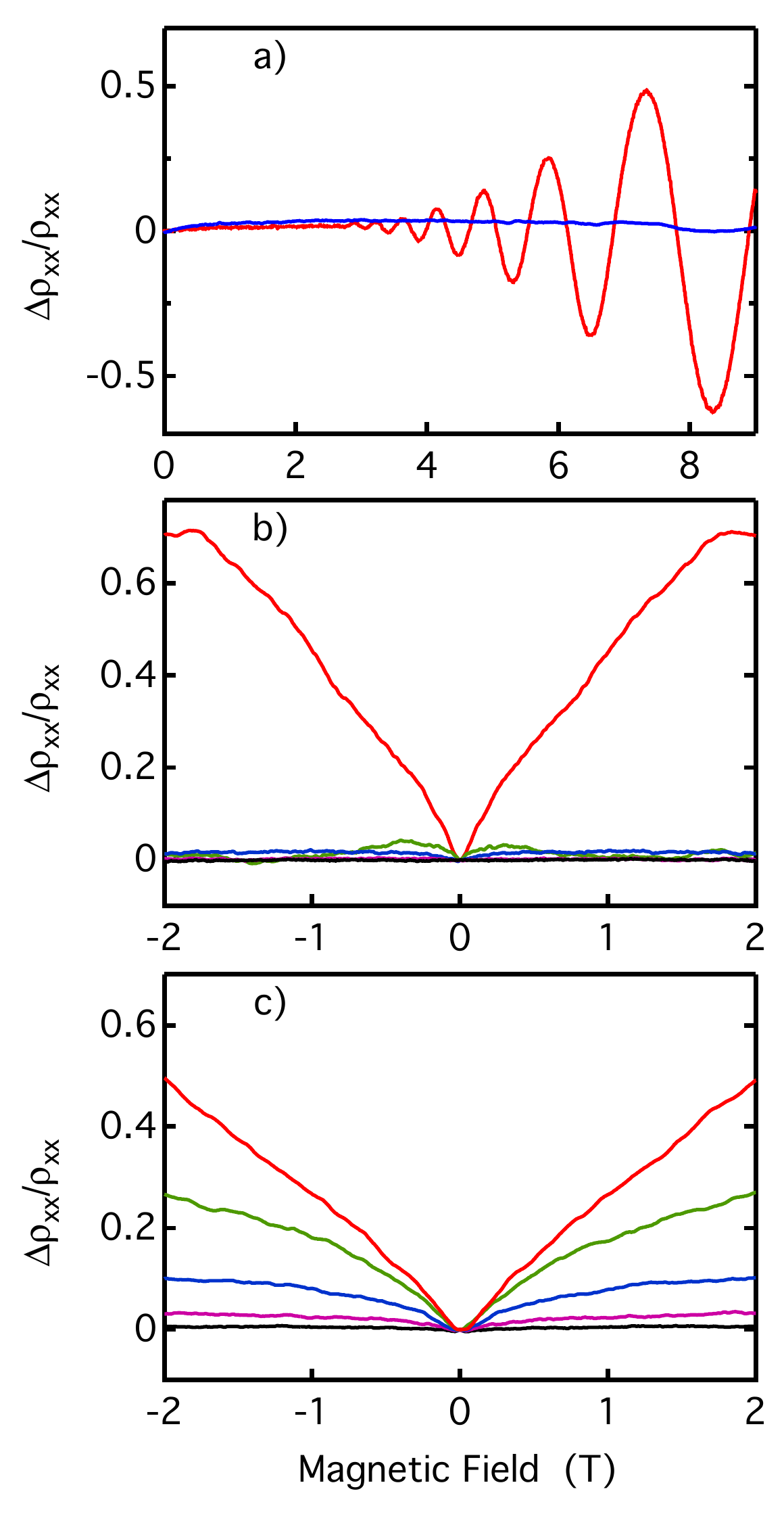}
\caption{(Color online) Effect of In adatoms on the magnetoresistance of graphene.  a) \drxx\ vs. $B$ at $n =-2.8 \times 10^{12}$ cm$^{-2}$.  Red trace corresponds to clean graphene sample, while for the blue trace approximately $2.5 \times 10^{12}$ cm$^{-2}$ In adatoms are present. b) and c): Magnetoresistance at various free carrier densities in the clean and In-decorated sample, respectively.  Red, green, blue, magenta, and black traces: $n=0$, 0.7, 1.4, 2.8, and $5.6 \times 10^{12}$ cm$^{-2}$.  All data at $T = 12$ K. }
\end{figure}
For the clean graphene sample, the large magnetoresistance seen at $n=0$ disappears quickly as $n$ becomes finite.  The green, blue, magenta, and black traces in Fig. 4(b), corresponding to $n = 0.7$, 1.4, 2.8, and $5.6 \times 10^{12}$ cm$^{-2}$, respectively, reveal virtually no magnetoresistance (for $B \leq 2$ T) at these densities.  In contrast, for the In-decorated sample, the strong, quasi-linear magnetoresistance found at $n=0$ subsides only gradually with density.  Moreover, as Fig. 4(c) demonstrates, \drxx\ at finite $n$ exhibits a non-linear, saturating dependence on magnetic field.

These same basic magnetoresistance effects are observed in \drxx\ data from the narrow region of the graphene sample, even if some quantitative differences do appear.  In addition, as already suggested by the $B=0$ data in Figs. 1 and 2, we find that the magnetoresistance, at a given $|n|$, is essentially identical on the electron ($n>0$) and hole ($n<0$) sides of the Dirac point.

Strong, quasi-linear, positive magnetoresistance of graphene near the Dirac point has been reported previously\cite{cho08,ping14}.  It generally believed to be a result of the charge density inhomogeneities (electron and hole ``puddles'') which are known to exist near the $n=0$ charge neutrality point.  As the average density $|n|$ increases, the system eventually becomes unipolar, with the fractional density fluctation $\sigma_n/|n|$ falling with $|n|$ owing to enhanced screening.  It seems reasonable to expect the magnetoresistance to subside once $\sigma_n/|n|$ is sufficiently small and the graphene carriers become more Drude-like.  That we observe a strong positive magnetoresistance over a wider density range about $n=0$ in the In-decorated sample than in the clean sample further supports our conclusion that the In adatoms exacerbate the density inhomogeneities in the sample.

Some aspects of the positive magnetoresistance features observed near the Dirac point are captured by theories which treat transport in a system of electron and hole puddles within an effective medium approximation (EMA)\cite{guttal05,tiwari09,rossi09}.  For example, EMA models\cite{guttal05} which assume electron and hole puddles having the same carrier mobility $\mu$ predict a linear magnetoresistance at $n=0$, at least for magnetic fields $B>>\mu^{-1}$.  Similarly, the observed non-linear saturating behavior of \drxx\ with $B$ at finite $|n|$ also emerges from EMA calculations.  Although estimating the mobility of the electron and hole puddles presumably present at $n=0$ is problematic, we note that the linear magnetoresistance we observe near $n=0$ persists to magnetic fields considerably smaller than $\mu^{-1}$, if $\mu$ is taken to be any of the values ($\mu \sim 2000-6000$ cm$^2$/Vs) found in our sample at densities well away from the Dirac point.  Finally, we note that alternative theories\cite{kozlova12} of the linear magnetoresistance, based on multiple scattering off of poorly conducting regions in an inhomogeneous conductor, have also recently appeared.    

\subsection{Quantum interference effects}
At magnetic fields below about $B \sim 100$ mT the magnetoresistance of both the clean and indium-decorated sample shows clear signs of quantum interference effects, including both weak localization and universal conductance fluctuations.  Figure 5 a) shows the change in the magnetoconductance $\Delta \sigma = -\Delta \rho / \rho^2$ around $B=0$, at various temperatures, for the clean graphene sample at a free carrier density of $n\approx 5 \times 10^{12}$ cm$^{-2}$.   These data exhibit a cusp-like minimum in $\Delta \sigma$ at  $B=0$, of depth comparable to $e^2/h$. As the temperature is increased, the minimum broadens and weakens.  In addition, the fluctuations in the conductance, which are roughly symmetric in magnetic field and quite prominent at the lowest temperatures, subside entirely by $T = 50$ K.  These features are clearly reminiscent of the signatures of weak localization and universal conductance fluctuations seen in ordinary disordered thin metal films\cite{lee85a,lee85b}.

\begin{figure}
\includegraphics[width=1 \columnwidth]{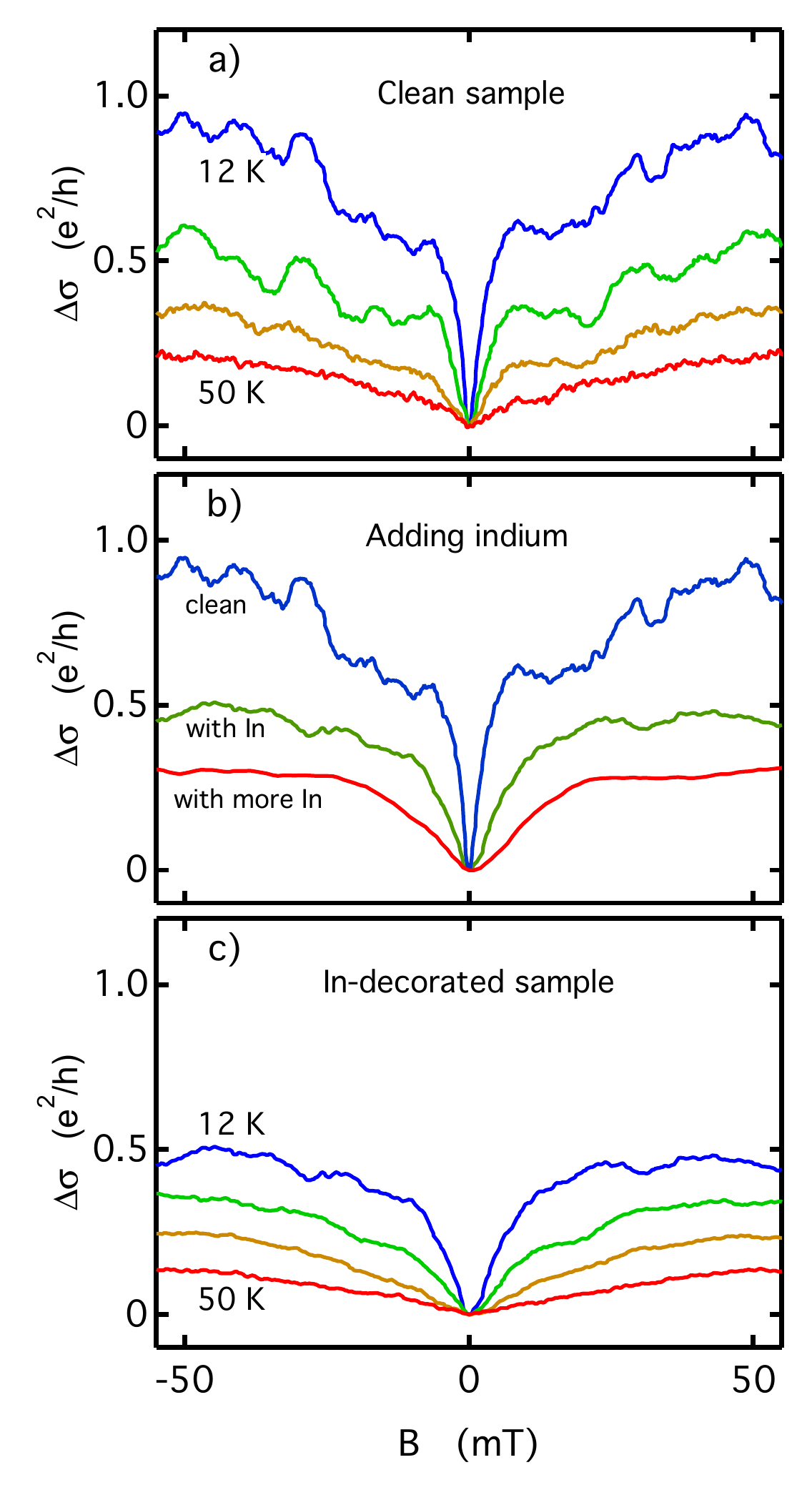}
\caption{(Color online) Low field magnetoconductance observed in the narrow region of the sample at a free carrier density of $n=5.5 \times 10^{12}$ cm$^{-2}$.  a) Clean sample at $T=12$, 20, 33, and 50 K.  b) Effect of In deposition on $\Delta \sigma$ at $T = 12$ K. Blue: clean sample. Green (Red): After deposition of approximately 2.5 (4.5) $\times 10^{12}$ cm$^{-2}$ In adatoms.  c) Magnetoconductance at $T = 12$, 20, 33 and 50 K of sample decorated with $2.5 \times 10^{12}$ cm$^{-2}$ In adatoms.}
\end{figure}
Figure 5 b) shows the effect of indium adatoms on the magnetoconductance at $T=12$ K.  The top trace corresponds to the clean graphene sample, while for the middle and lower trace the sample is decorated with approximately 2.5 and $4.5\times 10^{12}$ cm$^{-2}$ In adatoms, respectively.  In each case the gate voltage $V_g$ was adjusted to yield the same free carrier density, $n\approx 5\times 10^{12}$ cm$^{-2}$; this relatively high density was chosen in order that the complicating effects of carrier density inhomogeneity, discussed in section III-B, are minimized.  As the figure makes plain, the In adatoms reduce the magnitude and broaden the width of the cusp-like minimum in $\Delta \sigma$ around $B = 0$.  The conductance fluctuations, so readily apparent in the clean sample at low temperatures, are almost entirely suppressed by the In adatoms.  As explained below, these various changes in the magnetoconductance can be understood as consequences of the reduced mobility of the graphene carriers in the In-decorated sample.

Finally, Fig. 5 c) displays the temperature dependence of the magnetoconductance of the graphene sample decorated with approximately $2.5 \times 10^{12}$ cm$^{-2}$ indium adatoms.  Just as for the clean sample, increasing the temperature weakens and broadens the cusp-like minimum in $\Delta \sigma$.  Here again the free carrier density is $n \approx 5\times 10^{12}$ cm$^{-2}$.   At this density the mobility is found to be $\mu \approx 5400$ and 2300 cm$^2$/Vs in the clean and In-decorated sample, respectively.

The data shown in Fig. 5 were obtained from the narrow region of the Hall bar.  Data taken from the wide region of the Hall bar display the same cusp-like minimum in $\Delta \sigma$ although it is not as deep as that seen in the narrow region.  Possible explanations for this difference are discussed below.

The theory of weak localization in graphene\cite{mccann06,aleiner06,mccann12} is more intricate than the corresponding theory for ordinary metal films\cite{altshuler80,hikami80,maekawa81}.  A description including only two time scales, an elastic scattering time $\tau_{\mu}$ and a dephasing time $\tau_{\phi}$ is insufficient to capture the necessary physics.  In graphene, intervalley scattering, sublattice symmetry-breaking processes, trigonal warping of the Dirac cones, and weak spin-orbit effects all must be included in a complete theory.  Indeed, it is generally believed that intervalley scattering explains why weak {\it localization} is observed in graphene instead of the weak {\it anti}-localization originally anticipated to arise from the chirality-induced absence of intravalley backscattering.

In this paper we use a simplified version of the theory of McCann {\it et al} \cite{mccann06}, including only the elastic scattering time $\tau_{\mu}$, the dephasing time $\tau_{\phi}$, and an intervalley scattering time $\tau_i$.  We are thus ignoring sublattice symmetry-breaking processes, band warping, and spin-orbit effects.   In this simplified approach, the change in the magnetoconductance, $\Delta\sigma=\sigma(B)-\sigma(0)$, is given by
\begin{gather}
\Delta \sigma=\frac{e^2}{\pi h}[F(\frac{B}{B_{\phi}})-F(\frac{B}{B_{\phi}+2B_i})-2F(\frac{B}{B_{\phi}+B_i})]\nonumber\\
F(z)={\rm ln}(z)+\psi(\frac12+\frac1z), ~~B_{\phi,i}=\frac{\hbar}{4De}\tau^{-1}_{\phi,i}
\end{gather}
Here $D=v_F^2\tau_{\mu}/2$ is the diffusivity and $\psi$ the digamma function. Numerical estimates of $\tau_{\phi}$ and $\tau_i$ are obtained by fitting our data to Eq. 2 over the magnetic field range $|B|\le 25$ mT.  With this relatively narrow field window, the fits obviously emphasize the cusp-like feature in $\Delta \sigma$ at $B = 0$ at the expense of its behavior at larger fields.  This approach is justified, we believe, by several considerations.  First, there are observed contributions to the magneto-resistance at intermediate fields which are not captured by Eq. 2, notably those due carrier density inhomogeneity and universal conductance fluctuations.  Second, the applicability of Eq. 2 is limited to fields small enough that the elastic mean free path $\ell_{\mu}=v_F\tau_{\mu}$ is much less than magnetic length\cite{lee85a,hikami80} $\ell_0=(\hbar/eB)^{1/2}$.  For the clean graphene sample, at a free carrier concentration of $5\times10^{12}$ cm$^{-2}$, $\ell_{\mu}=\ell_0$ already at $B \approx 25$ mT.  Finally, we find that the inclusion of additional fitting parameters (e.g. sublattice symmetry-breaking times) leads to multiple chi-squared minima in the fitting procedure and large uncertainties in some of the extracted scattering times.

\begin{figure}[t]
\includegraphics[width=1 \columnwidth]{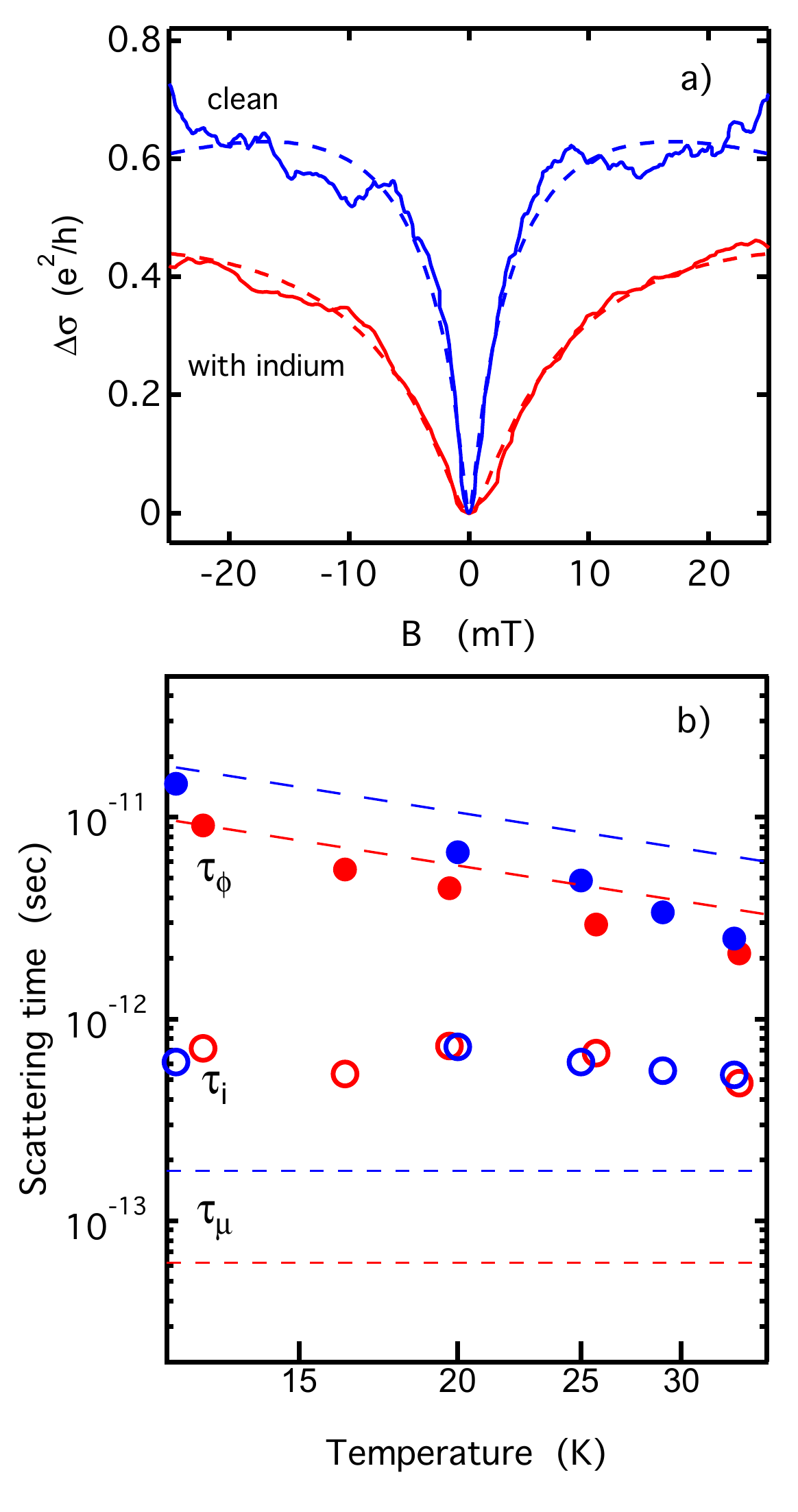}
\caption{(Color online) Results of weak localization data analysis at $n\approx 5\times10^{12}$ cm$^{-2}$.  a) Examples of fits (dashed) of Eq. 2 to data (solid) at $T=12$ K.  Blue:  clean graphene sample; Red: After deposition of approximately $2.5 \times 10^{12}$ cm$^{-2}$ indium adatoms. b) Fitted values of $\tau_{\phi}$ (solid circles) and $\tau_i$ (open circles) for the clean (blue) and In-decorated (red) graphene sample.  Diagonal (horizontal) dashed lines: theory estimates of $\tau_{\phi}$ ($\tau_{\mu}$) in graphene at mobilities 5400 (blue) and 2300 (blue) cm$^2$/Vs.} 
\end{figure}
Figure 6 a) shows two examples of the fits of Eq. 2 to data obtained from the clean and In-decorated sample.  The fits are clearly good at very low magnetic field.  At higher fields, especially for the clean sample, the fit worsens. While this is partly a result of the stronger conductance fluctuations evident in the clean sample, it might also suggest that we have over-simplified the McCann theory in arriving at Eq. 2, or merely violated the $\ell_{\mu}<<\ell_0$ requirement too egregiously. 

Figure 6 b) presents the fitted values of the dephasing time $\tau_{\phi}$ and intervalley scattering time $\tau_i$ for the clean and In-decorated sample at  free carrier density $n\approx5 \times 10^{12}$ cm$^{-2}$.  The diagonal dashed lines are theoretical predictions of $\tau_{\phi}$ in a diffusive 2D conductor from the work of Altshuler, Aronov and Khmel'nitski\cite{altshuler82}:
\begin{equation}
\tau_{\phi}^{-1}=\frac{k_BT}{2\pi\hbar^2D\nu}{\rm ln}(\pi\hbar\nu D),
\end{equation}
with $D$ again the diffusivity and $\nu$ the density of states at the Fermi level.  The upper and lower diagonal lines represent Eq. 3 with $D$ and $\nu$ evaluated for graphene with mobilities of 5400 and 2300 cm$^2$/Vs, respectively, at $n=5\times10^{12}$ cm$^{-2}$.  These predictions are in reasonably good agreement with the fitted values of $\tau_{\phi}$.

The fitted values of $\tau_i$, the intervalley scattering time, are essentially temperature independent and statistically identical for the clean and In-decorated sample. This last observation is intriguing since it implies that the intervalley scattering length $\ell_i=(D \tau_i)^{1/2}$ is shorter, by about a factor of $(5400/2300)^{1/2}\approx1.5$, in the In-decorated sample. Thus some fraction, crudely of order 10\%, of the scatterings off of the indium adatoms appear to be $intervalley$ processes.  It seems at least conceivable that the adatoms weakly distort the graphene lattice in their immediate vicinity (in addition to providing a long range Coulomb scattering potential) and thereby enable intervalley scattering.  Alternatively, this somewhat surprising result could be an artifact of our simplified weak localization analysis. 

The dephasing times $\tau_{\phi}$ found here lead to dephasing lengths $\ell_{\phi}=(D\tau_{\phi})^{1/2}$ ranging from $\ell_{\phi}\approx 1.0$ $\mu$m at $T=12$ K in the clean graphene sample to $\ell_{\phi}\approx 0.26$ $\mu$m at $T=33$ K in the In-decorated sample.  Since the width of the narrow region of the Hall bar is $W=1$ $\mu$m, the data described above may be in a cross-over regime from 2D to 1D localization.  This, in addition to the fact that boundary scattering is more important in the narrow region of the device than the wide region, may explain the why the weak localization signatures are stronger in the narrow region of the Hall bar. 

One motivation for depositing In adatoms onto graphene is that their strong spin-orbit interaction will convert graphene from a gapless semi-metal into a topological insulator ~\cite{weeks11,hu12}.  Weak localization is well-known to be a sensitive probe of spin-orbit effects in metals \cite{hikami80,maekawa81,bergmann82}.  In graphene, the theoretical situation is again more complicated than in ordinary metals\cite{mccann12}.  Although our simplified analysis omits spin-orbit scattering at the outset, it seems clear from the data that no weak anti-localization features analogous to those observed in metal films having strong spin-orbit scattering ({\it e.g.} Mg:Au \cite{bergmann82}) are seen.  This is perhaps not surprising given the very low In coverages ($\sim 0.1\%$) employed here.

\section{Conclusion}
Transport studies of graphene decorated with dilute concentrations of indium adatoms have been reported here.  Our results reveal that the In adatoms electron dope the graphene, as evidenced by gate voltage shifts of the Dirac.  The mobility of free carriers in the graphene is significantly reduced by the indium adatoms.  The near-linearity with density of the graphene conductivity after indium is deposited shows that the adatoms act primarily as long-range coulombic scatterers.  At the same time, our results reveal a pronounced broadening of the conductivity minimum and Hall coefficient at the Dirac point, demonstrating that the In adatoms increase the level of charge density inhomogeneity in the system. 

Analysis of the conductivity and Hall effect data via the theory of Adam {\it et al.} suggests that the In adatoms reside approximately $d \approx 0.8$-0.9 nm from the graphene surface.  This  seems surprisingly large given that the In readily dopes the graphene with electrons.   It is also in disagreement with recent density functional theory calculations which find $d\approx 0.24$ nm.  Resolving this discrepancy requires additional experimental work.

At intermediate magnetic fields a strong quasi-linear positive magnetoresistance is observed at the Dirac point.  This finding, which has been reported previously, has been attributed to the existence of electron and hole ``puddles'' at the (net) charge neutrality point.  While this effect subsides rapidly upon moving away from the Dirac point in the clean graphene sample, it persists to considerably higher density after indium adatoms have been deposited.  This further supports our conclusion that the In adatoms exacerbate the charge density inhomogeneity in the graphene.

Obvious signatures of weak localization and universal conductance fluctuations are seen at low magnetic field.  These signatures are modified when In adatoms are deposited in a way largely attributable to the reduced carrier mobility which results.  A simplified version of the theory of weak localization in graphene suffices to fit the magnetoconductance at very low fields and the extracted values of the dephasing times $\tau_{\phi}$ are in reasonable agreement with theory.  Plausible values of the intervalley scattering time $\tau_i$ are also obtained from the analysis, along with intriguing evidence that the In adatoms, while dominantly sources of long range intravalley Coulomb scattering, occasionally enable intervalley events as well.  No clear-cut evidence for spin-orbit effects is apparent in our data.  We note that similar conclusions have been reached in the work of Jia {\it et al.} \cite{jia15}.

\begin{acknowledgements}
We thank S. Adam, I. Aleiner, J. Alicea, S. Das Sarma, J. Hu, K. Kechedzhi, E. McCann, R. Mong, J. Pollanen, G. Refael, and R. Wu for helpful correspondence and discussions.  This work was supported by DOE grant FG02-99ER45766 and the Institute for Quantum Information and Matter, an NSF Physics Frontiers Center with support of the Gordon and Betty Moore Foundation through Grant No. GBMF1250.
\end{acknowledgements}

\end{document}